\documentclass[12pt]{article}
\usepackage{amssymb}
\usepackage{amsmath}
\usepackage{epsfig}
\usepackage{subfigure}
\newcommand{\beff}{$\beta_{eff}$}
\newcommand{\sav}{$\left< s \right>$}
\newcommand{\iav}{$\left< i \right>$}

\begin{document}

\title{ SIR model on a dynamical network and the endemic state
of an infectious disease}

\author{M.Dottori and G.Fabricius$^1$ \\ 
\small{Instituto de Investigaciones Fisicoqu\'{\i}micas Te\'oricas 
         y Aplicadas,} \\
\small{ Facultad de Ciencias Exactas, 
         Universidad Nacional de La Plata,}\\ 
\small{ cc 16, Suc. 4, 
         1900 La Plata, Argentina} }

\maketitle

\begin{abstract}
In this work we performed a numerical study
of an epidemic model that mimics the endemic state of
whooping cough in the pre-vaccine era. 
We considered a stochastic SIR model on 
dynamical networks that involve local and global contacts
among individuals and analyzed the influence 
of the network properties on the 
characterization of the quasi-stationary state. 
We computed probability density functions (PDF) 
for infected fraction of individuals and found that they are well 
fitted by gamma functions, excepted the tails of the distributions 
that are q-exponentials.  
We also computed the fluctuation power spectra of infective time 
series for different networks. We found that network effects can be 
partially absorbed by rescaling the rate of infective contacts of 
the model. 
An explicit relation between the effective transmission rate of the disease
and the 
correlation of susceptible individuals with their infective nearest neighbours 
was obtained. This relation quantifies the known screening of infective 
individuals
observed in these networks. We finally discuss the goodness and limitations 
of the SIR model with homogeneous mixing and parameters taken from 
epidemiological 
data to describe the dynamic behaviour observed in the networks studied.
\end{abstract}


\addvspace{15pt}
\small{$^1$ {\it e-mail:} fabricius@fisica.unlp.edu.ar (corresponding author)}
\newpage

\section{Introduction}
\label{intro}

Mathematical modelling of infectious diseases is
an interdisciplinary area of increasing interest.
Since the pioneering work of Kermack and McKendrick \cite{kermack},
mathematical modelling
has shown to be a powerful tool to understand
infectious disease transmission 
\cite{libroAM,altizer,lavine,nos-epidemics}.
The fact that the simple deterministic SIR model, where the
 population is divided
into susceptible (S), infective (I) and immune or recovered (R) individuals, 
is able to predict quite well  the outbreak periods for
a large series of infectious diseases in many countries
around the world suggests that it 
captures the essentials of transmission dynamics \cite{libroAM}. 
Lately, models gained  an increasing level of complexity, 
including age structure and particular features of the given disease, 
with the aim of being more realistic and predictive
\cite{libroAM,Hethcote99, vanboven, nos-adol-boost}. 
Due to the
complexities of these models, 
stochastic effects and the network structure of contacts are rarely considered. 
However, when considered (in much more simple models) stochastic effects 
have shown to play an important role in understanding the data 
of reported cases of measles and pertussis \cite{science-rohani-99}. 
More recently, the effect of the network structure of contacts on the 
disease transmission has been considered in several publications, 
always in models with very few epidemiological classes 
\cite{redes-may,kuperman,redes-keeling,roypascual}.
      In particular, in Refs. \cite{verdasca, simoes} the authors 
studied the spatial correlations and stochastic fluctuations using SIR
 and SEIR models on two-dimensional Watts-Strogatz-type dynamical networks.
 Sim\"oes {\it et al.} \cite{simoes} focussed on  the
 power spectrum of the fraction of infective time series and performed an 
extensive study of parameter space for the SIR model.
They found that spatial correlations and the deterministic recovery of infection
 increase the amplitude and coherence of the resonant stochastic fluctuations 
and studied the dependence of such changes on model parameters.

In the present work, we use the SIR stochastic model
on two-dimesional Watts-Strogatz-type dynamical networks
and perform an intensive study of the model for
parameters corresponding to pertussis disease 
(whooping cough) in the pre-vaccine era.
We carry out simulations for different parameters characterizing the network
properties and study their influence on the behaviour
of the quasi-stationary state of the system
corresponding to the endemic disease with
periodic outbreaks.
The purpose of our work is twofold.
On the one hand, we assume the disease is propagating in a place
with a well-mixed population, and then, for some
reason, the network of contacts changes and becomes more local.
We analyzed the consequences of this change in disease transmission.
On the other hand, we focus on a
methodological point. Suppose the disease is 
propagating in a city where local contacts are important,
but a SIR stochastical model disregarding network
structure is used. How bad is the description
of the problem in this case if you parametrize the model
to available epidemiological data?

\section{Model and simulations}
We consider a stochastic SIR model on Watts-Strogatz dynamic-type networks 
as the ones studied by Verdasca {\it et al.} \cite{verdasca}  
and  Sim\"oes {\it et al.} \cite{simoes}.  
Our setup of the model is summarized below. 
First, we describe
the underlying SIR deterministic model with births and deaths.

\subsection{Deterministic SIR model}

In this model it is assumed that individuals are in one of the following
three epidemiological classes: susceptible, infected or 
   recovered. Individuals are born in a susceptible class 
at a rate $\mu$ and they
remain there until they become infected by contact with
an infected individual. Infected individuals recover from infection
entering the recovered class at a rate $\gamma$.
The dynamic variables of the model are the fractions
of the population in each epidemiological class, and they obey
the following set of non-linear coupled differential equations:
\begin{flalign}
& \frac{ds}{dt}  =  -\beta s i + \mu  - \mu s  \nonumber     \\
\label{eqsdeterm}
& \frac{di}{dt}  =  \beta s i - \gamma i - \mu i            \\
& \frac{dr}{dt}  =   \gamma i - \mu r             \nonumber 
\end{flalign}
where $s$, $i$ and $r$ are the fractions of people in susceptible, 
infected and recovered classes respectively. 
In this model the term $\beta s i$ represents the incidence 
per individual of the disease
(the rate at which susceptible individuals become infected) and contains the key
approximation of the model: uniform mixing. Parameter
$\beta$ is the rate of infective contacts (which are the contacts such
that if one individual is infected and the other susceptible, the latter
will become infected). So, in the uniform mixing approximation
it is assumed that all susceptible individuals become infected 
at the same rate: $\beta i$.
In this model the death rate is assumed equal for people 
in the three epidemiological
classes and also equal to the
birth rate, $\mu$, in order to keep the population constant.      

The set of differential equations \ref{eqsdeterm} determines the dynamic
evolution 
of the system. For any initial conditions different from i=0 
(when the system goes towards the fixed point: $s=1$, $i=0$, $r=0$), 
the system asymptotically reaches 
the stationary state: $s=s^*=(\gamma + \mu)/\beta$, $i=i^*=
\mu/(\gamma + \mu) (1-s^*)$, $r^*=1-s^*-i^*$.

\subsection{Stochastic model on the network}

We consider N individuals on a squared lattice (LxL=N) 
with periodic boundary conditions. At each site
there is an 
individual that may be in one of the 3 epidemiological states: S (suscepible), 
I (infected) or R (recovered). 
The state of an individual at site $j$ is a stochastic 
variable of the model, $X_j$, that may change through the following processes:

\begin{flalign}
& infection:  & S \rightarrow I &&&&&&& \nonumber \\
& recovering: & I \rightarrow R  &&&&&&& \nonumber \\
& death \ and \ birth: & S \rightarrow S \nonumber &&&&&&& \\
&                  & I \rightarrow S \nonumber &&&&&&& \\
&                  & R \rightarrow S \nonumber &&&&&&& 
\end{flalign}

We assume   that when an individual dies at a site, another individual is born 
at the same time at this site in order
to keep every site with one individual during the simulation. As we suppose 
Markovian processes, the dynamics of the system is controlled by the knowledge
of probability transition rates at each time. 
Deaths and births
are independent of the individual state and occur at the same 
probability rate $\mu$.
To account for infections we consider a dynamic type of 
Watts-Strogatz network \cite{watts-strogatz}.
We assume that, at a given time, an individual at site $j$
has contact with a randomly chosen individual in the network 
with probability rate  $p \beta$, and with one of their $k$ nearest
neighbours with probability rate $ (1-p) \beta$. 
If the individual at site $j$ is susceptible, and the contacted individual
is infected, then the individual at site $j$ will become infected. 
Local contacts of an individual 
represent the contacts with known people
(in the circle of their stable relations)
while global random contacts represent
people met by chance (for example, on a bus, shopping, etc). 
Actually, here the word ``contact'' is restricted to ``infective contacts'',
in the sense discussed above for the SIR model.
The model used in the present work allows changing the degree of 
``locality" of the network by changing the value of parameter $p$. 
In particular, for $p=0$ an individual only has contacts with 
their $k$ nearest neighbours while for $p=1$ an individual 
may contact any other individual in the lattice with the same 
probability as in the classical stochastic SIR model (uniform mixing). 
This is an important difference with the standard (static) 
Watts-Strogatz setting \cite{watts-strogatz} 
where the case $p=1$ corresponds 
to a random network where an individual has a fixed 
number $k$ ($<<$N) of random contacts.
Recovery from infection is the same in every site and
occurs at a probability rate $\gamma$. This assumption
gives exponentially distributed recovery times,
which is a reasonable approximation for pertussis \cite{rohanigamma}.

In summary, 
the probability transition rates for 
infection, recovery and death-birth processes 
at site ``$j$" are
\begin{flalign}
&a_{inf}^j=\left[ p \ \beta \ i + (1-p) \beta \ 
            \frac{1}{k} \sum_{j'\in \nu_j} 
            \delta _{X_{j'},I} \right] \delta _{X_{j},S} \nonumber \\  
\label{aj}
&a_{rec}^j=\gamma \delta _{X_{j},I} \\
&a_{d-b}^j=\mu  \nonumber
\end{flalign}
where $\delta_{A B}$ is one if states $A$ and $B$ are the same, 
and zero if not. 
$X_j$ is the state of the individual at site $j$, and
$j'$ in the sum runs over the $k$ neighbours of site $j$ (we call 
this set of sites $\nu_j$).

The probability rate for infections, recoveries or death-birth processes
in the whole system is obtained summing over $j$ in Eqs. \ref{aj}
and gives
\begin{flalign}
\label{ainf}
& a_{inf}=  p \ \beta \ i \ s \ N 
+ (1-p) \beta \ n_{S I_\nu} N \\ 
\label{arec}
& a_{rec} = \gamma i N \\
\label{adb}
&a_{d-b}= \mu N
\end{flalign}
where $i$ and $s$ are the fractions of infected and susceptible individuals
in the system, and $n_{S I_\nu}$ is given by
\begin{flalign}
&n_{S I_\nu} =  \frac{1}{N} \sum_{j=1}^N 
            \frac{1}{k} \sum_{j'\in \nu_j} 
            \delta _{X_{j},S} \ \delta _{X_{j'},I}   
\label{nsinu}
\end{flalign}

In the case $p=1$ and $N \rightarrow \infty$, for $i$ and $s$ 
the stochastic model gives 
the same dynamics as that of the deterministic Eqs.
 \ref{eqsdeterm}. For $p<1$ the dynamics will be affected
by the correlation among susceptible and infected neighbours
explicitely included in the term: $n_{S I_\nu}$.

\subsection{Simulation algorithm}

The state of the system at a given time $t^\alpha$ is specified by the
knowledge of the $N$ variables $X^\alpha_j$: 
${\bf X^\alpha} = (X^\alpha _1, X^\alpha _2, ..., X^\alpha _N)$.
We perform stochastic simulations using Gillespie algorithm 
\cite{gillespie}. 
This is an exact algorithm that generates a Markov chain
for the master equation that could be constructed 
from the probability rates given in Eqs. \ref{aj}.
The algorithm gives a sequence of times $t^1$, $t^2$, ... and
the corresponding states ${\bf X^1}$, ${\bf X^2}$, 
..., where two consecutive states
differ by a single process that occurs at a given site. 
The process and the time when it takes
place are generated from simple rules and two random numbers
(see Ref. \cite{gillespie} for details).
As the probabiliy rates are functions of the stochastic variables of the
model, they are changed at each step of the simulation.

\subsection{Average computation}
For each system to be studied we generate a set of $M$ Markov chains,
from specified initial conditions, and use different sets of random numbers,
checking that each trajectory  survives 
at least a time $t_{run}$.
We denote the state of the system corresponding to 
Markov chain {\bf m} at time $t^\alpha$: ${\bf X_m ^\alpha}$.
To compute the average of an observable $A \left[ {\bf X_m ^\alpha} \right]$
over the $M$ samples at a given time, we take into account
 that the set of discrete
times $t^1$, $t^2$, ..., $t^\alpha$,... will be different for each one
of the {\bf m}=1, ...,$M$ trajectories.
So, we define
\begin{equation}
\left<  A (t) \right>=\frac{1}{M} \sum_{m=1}^M A_m(t), \ \
A_m(t)=\left(A \left[ {\bf X_m ^\alpha} \right] + 
             A \left[ {\bf X_m ^{\alpha+1}} \right] \right) /2
\end{equation}
with $t^\alpha < t <t^{\alpha+1}$.
As $t^{\alpha+1}-t^\alpha$ is the time interval between 2 single processes,
it is much shorter than the time taken by the whole 
system to undergo a detectable change. 

Similarly, we define the time correlation of an observable
at two diferent times
\begin{equation}
\left<  A (t') A(t'') \right>=\frac{1}{M} \sum_{m=1}^M A_m(t) A_m(t')
\end{equation}

The main observable studied in the present work is the fraction
of infected individuals
\begin{equation}
i \left[ {\bf X_m ^\alpha} \right]=\frac{1}{N} \sum_{j=1}^N 
       \delta _{ {\bf X_m^\alpha}_j , I}
\end{equation}

\section{Results}
We perform simulations using the algorithm described in sections
2.2 and 2.3 for L=800 (which corresponds to a city of
N=640,000 inhabitants) and consider diferent networks corresponding
to the cases 
$k$= 4, 8 and 12, including up to first, second and third 
neighbours respectively, and $p$ varying from 1.0 to 0.2.
We take $\mu=1/(50$ years), $\gamma=1/(21$ days) and $\beta=0.8$ 1/day, 
which are standard parameters for SIR description 
of pertussis in pre-vaccine era \cite{rozhnova}. When other
$\beta$ values are used in the simulations, it is mentioned explicitely.

\subsection{The quasi-stationary state}

\subsubsection{Definition and empirical assumption of its existence}
The purpose of this work is to simulate the endemic state of a disease that should be represented by a stationary state of the model. As we mentioned above in the case 
$p=1$, $N \rightarrow \infty$,
the dynamics of the system follows Eqs. \ref{eqsdeterm}
and so, for every initial condition with $i \neq 0$
the stationary state $(s^*,i^*)$ is reached.
However, for any finite value of $N$ the only stationary state
of the system is $s=1, i=0$ because sooner or later a 
fluctuation will make the number of infected individuals zero,
and there is no process that produces infected individuals
if there is none in the system. 
Nevertheless, for N large enough, the system may fluctuate 
for a long time around a quasi-stationary state (QSS) before extinction.
The definition and properties of such a state have been addressed
in other contexts from a mathematical point of view 
\cite{daroch-seneta,nasell}
or with empirical approaches \cite{dickman,zheng}.
In the present work an empirical strategy is developed.
There are
two points  to be considered in order that the system reaches 
a QSS and remains there long enough to be studied.
For $p<1$ the system develops correlations and there is a
time needed to arrive at this correlated QSS 
that depends  on the initial conditions.
Moreover, the size of the fluctuations 
(and thereby the probability of extinctions) as well as $N$,
depend  on the parameters
that define network properties: $p$ and $k$.
Then, the time  window
where a QSS of the system
could be defined has to be determined with some care.

{\it Estimation of extinction times}

\begin{figure}[ht]
\subfigure[ ] {
\includegraphics[width=7.8cm,height=5.5cm]{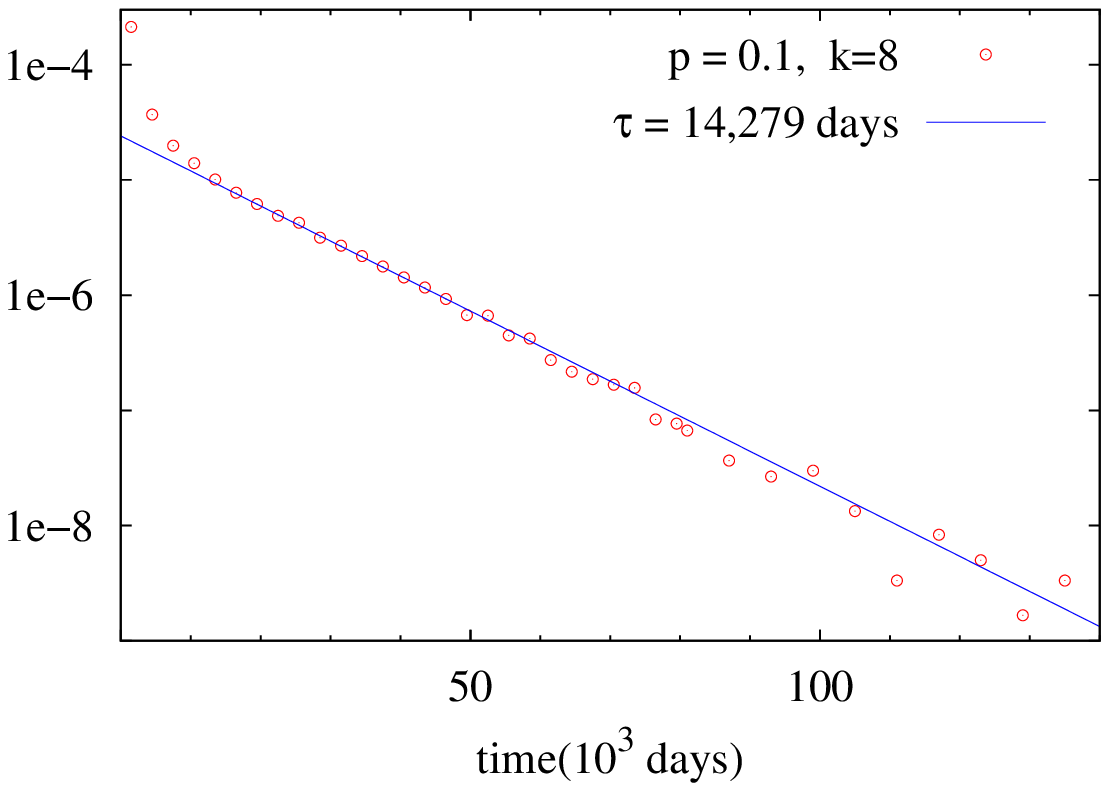}}
\subfigure[ ] {
\includegraphics[width=7.8cm,height=5.5cm]{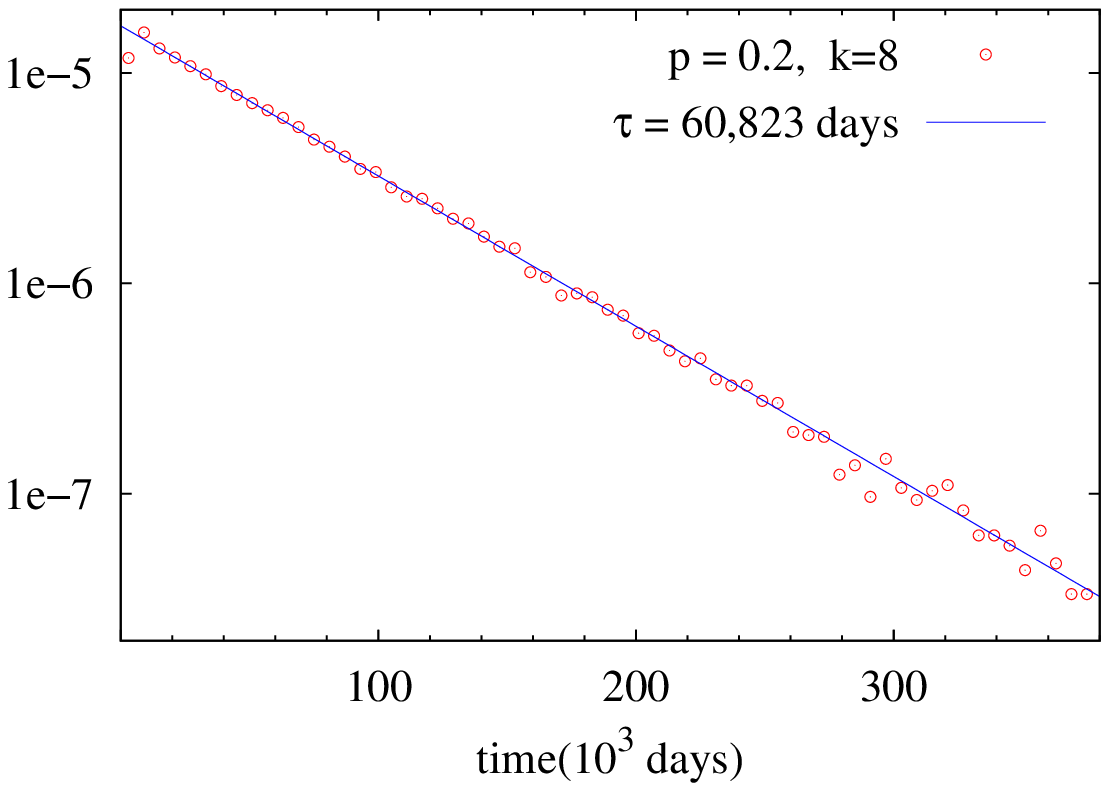}}
\subfigure[ ] {
\includegraphics[width=7.8cm,height=5.5cm]{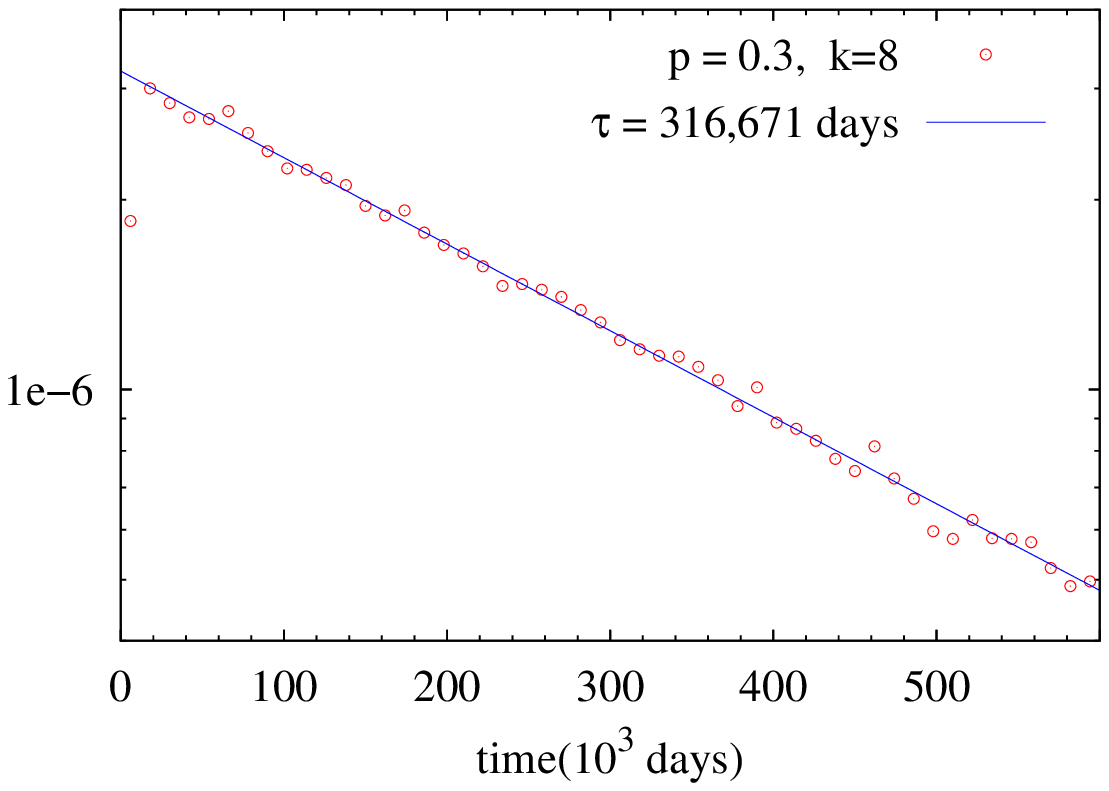}}
\subfigure[ ] {
\includegraphics[width=7.8cm,height=5.5cm]{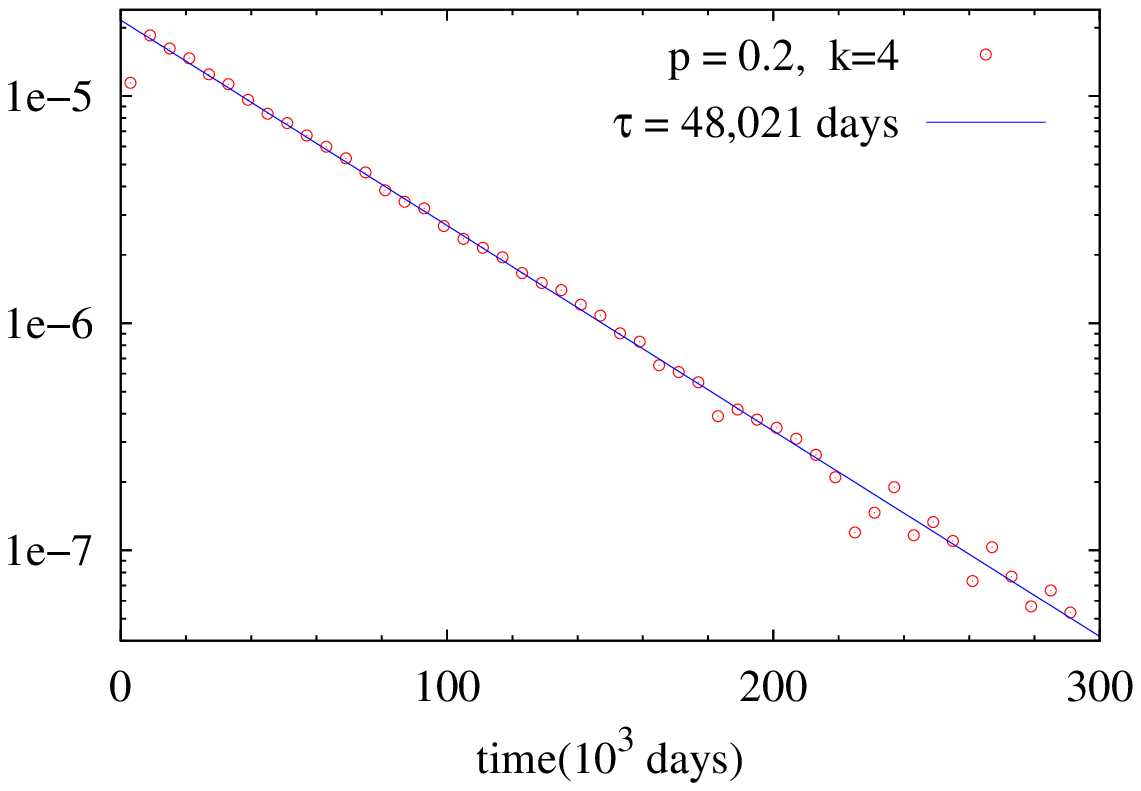}}
\caption{\label{dlt} Distribution of living times for 
different networks.
The points indicate the fraction of the samples considered
extinguished at a given time per unit of time.
The continuous lines 
are exponential fits $C \exp(-t/ \tau)$ to the data.
The fitted value of $\tau$ is shown in each figure.
}
\end{figure}

We first calculate the distribution of living times
in order to estimate the typical time that the  system
survives until extintion. We proceed as follows: 
(i) for each ($k,p$) network we perform several runs and, from
those that survive long enough, we obtain approximate values
for the average number of susceptible, infected and recovered 
individuals,
(ii) as an initial condition for our study, 
     in the network sites
     we randomly distribute
     a number of S, I and R individuals as obtained in (i), 
(iii) we generate M=50,000 different samples 
      as in (ii) (100,000 for $p=0.1$)
 and    let each sample evolve until extinction,
(iv) we compute the fraction of the samples that extinguished
in interval $(t, t+\Delta t)$.
The results are shown in Fig.\ref{dlt}. 
In all cases considered exponentials fit quite well the results of simulations. 
For the SIR stochastic model I. N$\mathring{{\rm a}}$ssel \cite{nasell} 
has proved that extinction times have 
an exponential distribution as we found here numerically 
for a more complex model and other initial conditions.
The typical time of extinction decreases sharply with $p$.
It is reduced to $1/20$ of its value in going from 
$p=0.3$ to $p=0.1$ for $k=8$.
Extinction times also decreases when lowering $k$ but
the effect is weaker, for example, they
undergo only a 20\% reduction
when the neighbours are reduced from $k=8$ to $k=4$, for $p=0.2$.

{\it Estimation of equilibration times}

\begin{figure}[ht]
\includegraphics{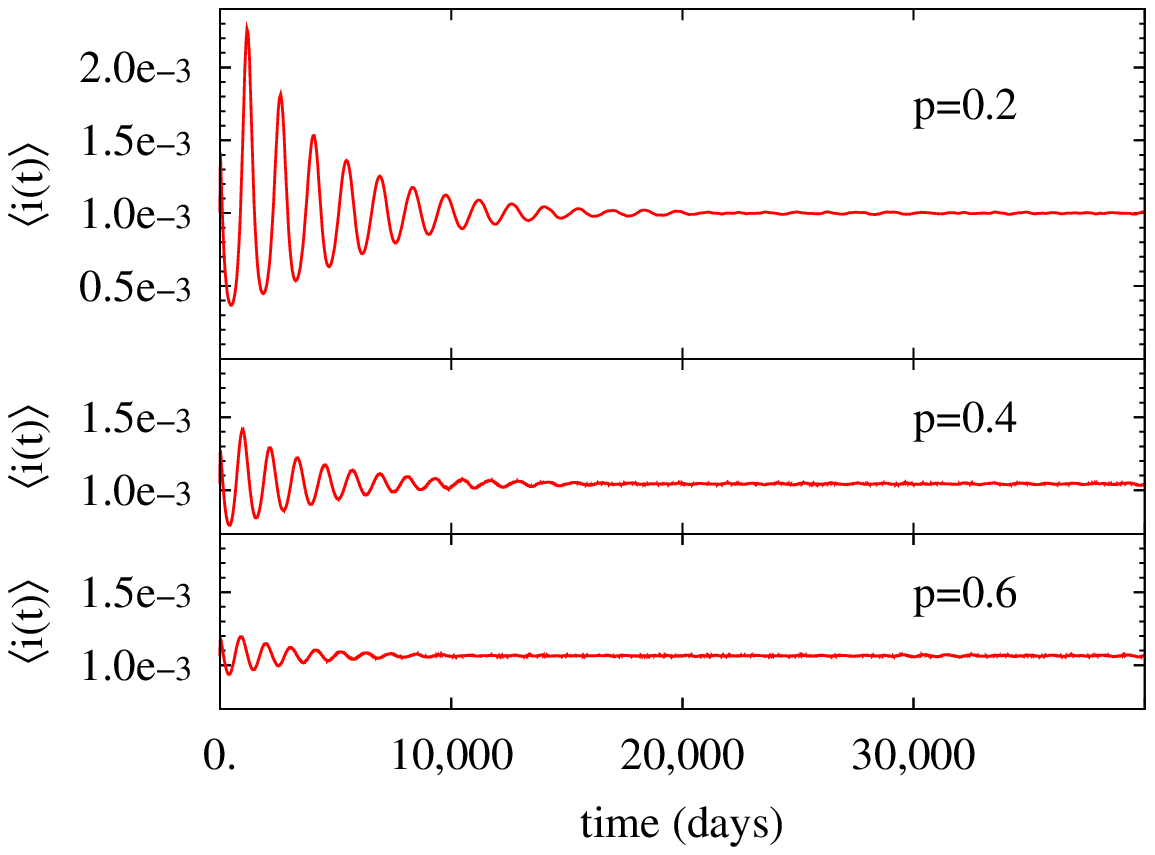}
\caption{\label{equil}  Average fraction of infected individuals
$\left< i(t) \right>$ 
as a function of time
for networks with $k=8$ and different $p$-values.
The average is performed over $M$=20,000 samples for each case.
}
\end{figure}

Figure \ref{equil} shows the time evolution of $\left< i(t) \right>$
computed averaging over $M$=20,000 samples that survive at least
a time $t_{run}=40,000$ days. The initial conditions for the simulations 
are generated as in (ii).
After a time that is longer for lower $p$, $\left< i(t) \right>$
remains around a constant value that we 
take as the stationary value of $\left< i(t) \right>$  
and denote it: $\left< i \right>$.
For $t>20,000$ days, the curves of Fig.\ref{equil}
verify $|\left< i(t) \right>- \left< i\right>|<\epsilon$=0.00001,
where $\left< i\right>$ is 0.001, 0.00104 and 0.00106 for $p$=0.2,
0.4 and 0.6, respectively.  
As the fraction of infective individuals in the system is the more fluctuating
magnitude we have studied, we assume its constancy in time is a sufficient 
condition to define the quasi-stationary state of the system.
The longer time needed to arrive at QSS as $p$ decreases is related
to the increasing time needed for the network to establish local
correlations. In particular, for the case $p$=1 (not shown in the figure)
the system is at QSS from t=0.
Taking  the time window $(t_a, t_b)$, 
$t_a$=20,000 days, $t_b=$40,000 days,
the 20,000 samples that survive 40,000 days for $p$=0.2 represent 
a significant 52\% of the generated samples.
Conversely, in the case $p$=0.1 for $t=$20,000, when QSS 
has not been reached yet, only 8\% of the samples survive. 
As in this work we are interested in the study and characterization 
of the endemic state of the disease, we haven't considered networks 
where the probability of establishment and survival of the steady 
state is very low.
In the present work we consider values of $p\geq 0.2$.

In summary, for each network ($k$, $p$) considered we obtained
a number $M$=20,000 of Markov chains generated 
from independent samples of the system
that survive at least a time  $t_b=$40,000 days.
In the  time window 
$(t_a, t_b)$
the magnitudes of interest
remain stationary within an acceptable precision ($\epsilon$) 
when averaged over the $M$ samples. From this empirical fact 
we assume the existence of a quasi-stationary state and that 
each one of the $M$ trajectories 
${\bf X_m ^\alpha}$ ({\bf m}=1,..M) represents
a possible time evolution of the disease in the endemic state
($t_a<t^\alpha<t_b$). 
The stationary value of an observable average $\left< A(t) \right>$
will be denoted $\left< A \right>$.
In particular, for the fraction of infected or susceptible 
individuals in the system
\begin{equation}
\left<  i (t) \right>=\left<  i \right> , \
\left<  s (t) \right>=\left<  s \right> , 
  ~ ~ {\rm for} \ t\in (t_a, t_b) 
\end{equation}
where the equality with the numerical values 
$\left< i \right>$  and 
$\left<  s \right>$ has to be understood to be valid within the precision 
$\epsilon$. 

\subsubsection{Relation between $ \left< s \right> $ 
                            and $\left< i \right> $}
In a given state ${\bf X^\alpha}$, 
the expected change in the fraction of infected individuals in 
the system is governed by the net probabilty rate
\begin{equation}
\left( a_{inf}[{\bf X^\alpha}] - a_{rec}[{\bf X^\alpha}] - i \ a_{d-b}[{\bf X^\alpha}] 
                          \right) /N
\end{equation}
but stationarity of $\left< i(t)\right>$ implies
\begin{equation}
\left< a_{inf} \right> -\left< a_{rec} \right>- \left< i\ a_{d-b}\right> =0
 \Rightarrow  \left<a_{inf}\right>/N = (\gamma+\mu) . \left< i \right>
\label{ist}
\end{equation}
where we have taken $a_{rec}=\gamma i$ and $a_{d-b}=\mu$
from Eqs. \ref{arec} and \ref{adb}. In the same way, 
the stationarity of $\left< s(t) \right>$ gives
\begin{equation}
- \left< a_{inf} \right> + \left< a_{d-b}\right> - \left< s \ a_{d-b} \right>  =0
\Rightarrow  \left< a_{inf} \right>/N =\mu . (1- \left< s \right>)
\label{sst}
\end{equation}
From Eqs. \ref{ist} and \ref{sst}
\begin{equation}
\left< i \right>=\frac{\mu}{\gamma+\mu}. (1-\left< s \right>)
\label{isst}
\end{equation}
i.e., $\left< s \right>$ and $\left< i \right>$ should be 
related by the same relation as that of $s^*$ and $i^*$ in 
  the deterministic model,
independently of the average rate of production of
infective individuals in the system, controlled by $\left< a_{inf} \right>$.  

\subsection{Stationary behaviour}

\begin{figure}[ht]
\includegraphics[width=9.8cm,height=7.5cm]{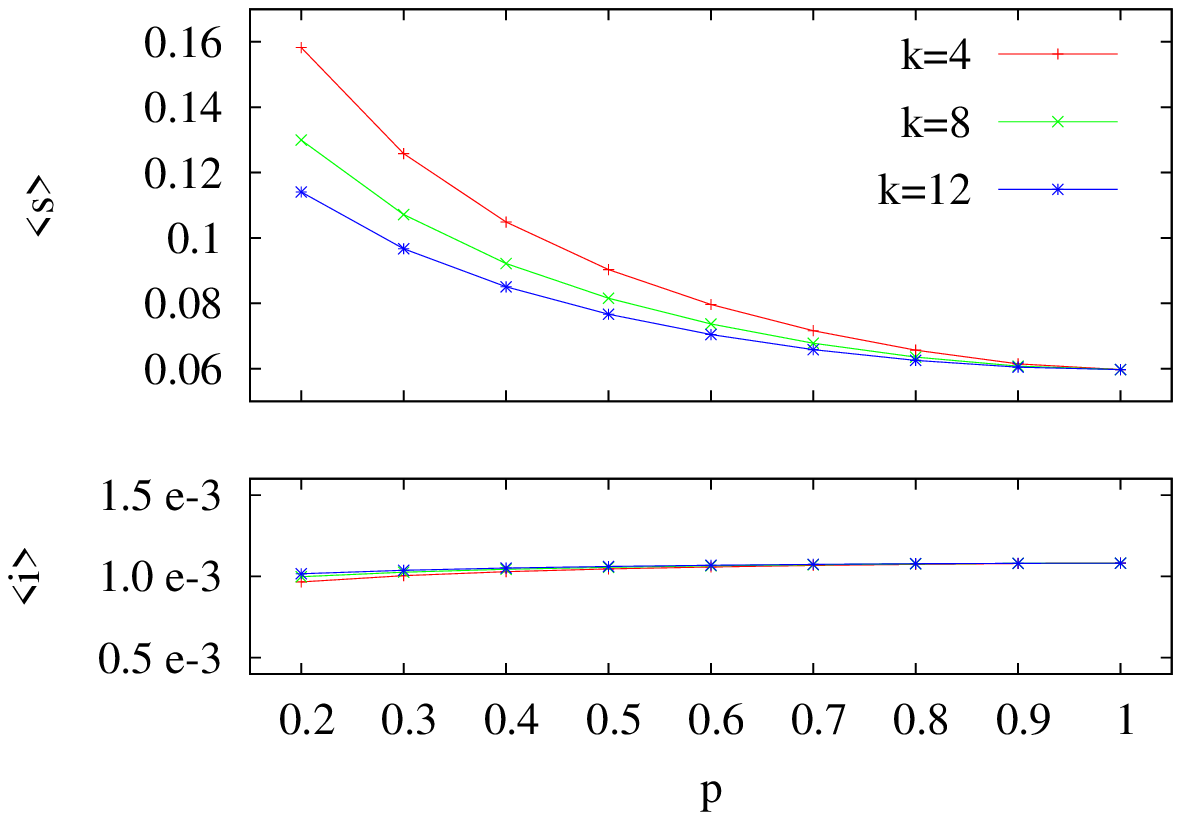}
\caption{\label{figsi} Stationary values of susceptible and
infected fractions in the network as a function of $p$ 
for a different number of neighbours $k$.
}
\end{figure}

In Fig. \ref{figsi} we show the stationary values of susceptible 
and infected fractions of individuals
for all the networks considered. 
While $\left< s \right>$ increases as  $p$ and $k$ decrease,
$\left< i \right>$ remains almost constant.
For the case 
p=0.2, k=4, $\left< s \right>$ is 2.6 times larger than for $p=1$
while $\left< i \right>$ 
decreases  only by 10\%.
The curves $\left< i \right>$ vs. $\left< s \right>$ (not shown) satisfy   
Eq. \ref{isst} within the precision of our simulations
for the 27 ($k$, $p$) networks considered. 
Given the $\left< s \right>$ dependence on  $p$ and $k$
shown in Fig.\ref{figsi}, a much weaker dependence of $\left< i \right>$
on network parameters is in fact predicted by Eq. \ref{isst},
since $\left< s \right>$ is always small compared with 1, and 
$\left< i \right>$ depends on network parameters through the factor
$(1-\left< s \right>)$.

The increase of $\left< s \right>$ when the locality of the network increases 
is a consequence of the decrease in effective transmission of 
the disease.
To quantify disease transmission in the 
network, we define the effective transmission rate
\begin{equation}
\beta_{eff}=\frac{\left< a_{inf}\right> /N}
                 {\left< i \right> \left< s \right>}
\label{defbeff}
\end{equation}
as the mean rate of infections per individual in the network, 
divided by the stationary values of fractions of infected and
susceptible individuals.
With this definition the average production of infections in the 
system is connected with \sav\ and \iav\ 
by the uniform mixing expression: 
\beff N\sav \iav, \beff\ playing the role of a ``global effective" 
contact rate in a problem where local and global contacts are present.

\begin{figure}[ht]
\includegraphics[width=9.8cm,height=7.5cm]{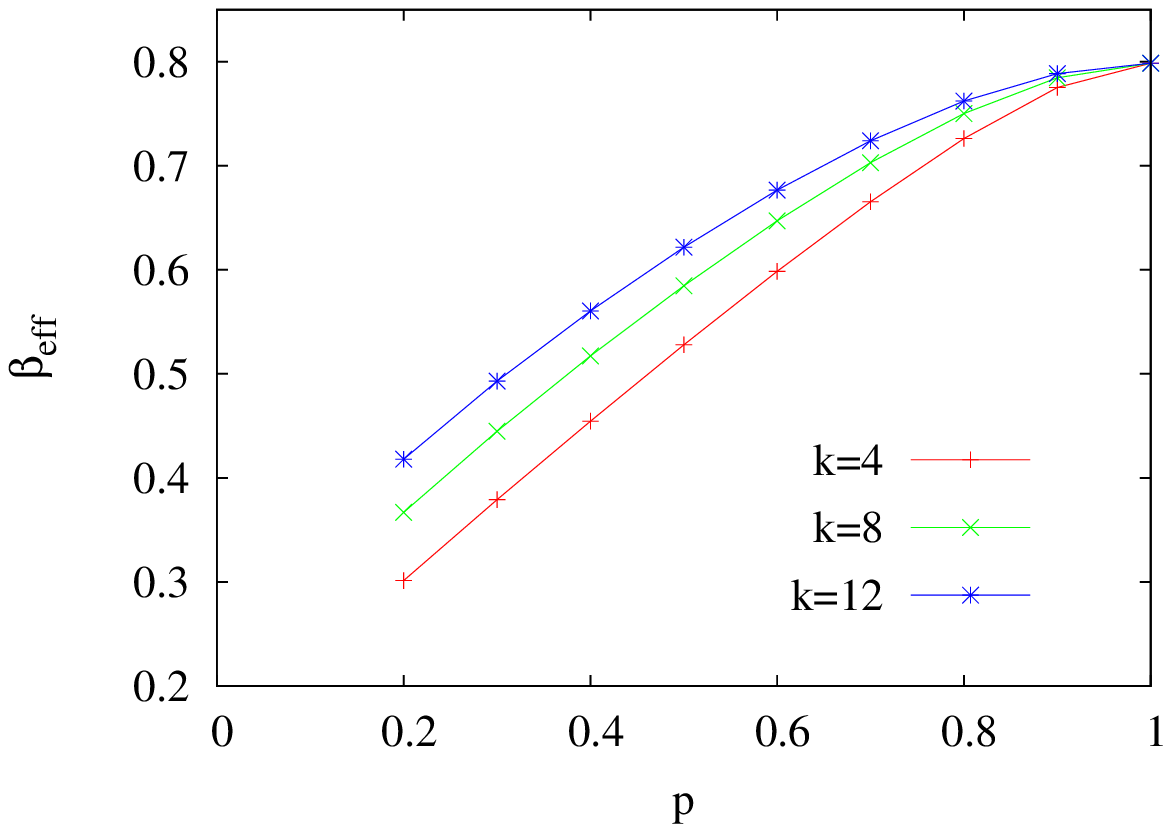}
\caption{\label{figbeff} Effective transmission rate
     $\beta_{eff}$ as a function of $p$ for a different
     number of neighbours $k$.
}
\end{figure}

In Fig. \ref{figbeff} we show $\beta_{eff}$ for all the 
networks considered. 
Transmission of the disease in the network decreases
with $p$ and $k$ as the network becomes more local. 
The drop in transmission with $p$ has been related to the 
clustering of susceptible and infected individuals that
is produced by local correlations \cite{verdasca, simoes}.
This clustering would reduce the probability of finding
$I-S$ neighbours and the local contribution to transmission.
In order to make the connection between 
disease transmission and local correlations quantitative, we express
 $\beta_{eff}$ as
\begin{equation}
\label{beffcorr}
\beta_{eff}=\beta + p \beta \frac{C_{s i}}{\left< i \right> \left< s \right>}
                  + (1-p) \beta \frac{C_{S I_\nu}}{\left< i \right> \left< s \right>}
\end{equation}
in terms of the correlation coefficients
\begin{equation}
\label{coefcorr}
C_{s i}    = \left< s i \right> - \left< i \right> \left< s \right>, \ \
C_{S I_\nu}= \left< n_{S I_\nu} \right> - \left< i \right> \left< s \right>
\end{equation}
where we have taken the average of Eq. \ref{ainf} 
to replace $\left< a_{inf}\right>$ 
in Eq. \ref{defbeff}.
Correlation $C_{s i}$ measures the average fluctuation
of the product of the fractions of susceptible and infected
individuals in the population, while $C_{S I_\nu}$ is a measure
of the local correlation between $S$ individuals with their $I$ 
neighbours. 
\begin{figure}[ht]
\subfigure[ ] {
\includegraphics[width=7.8cm,height=5.5cm]{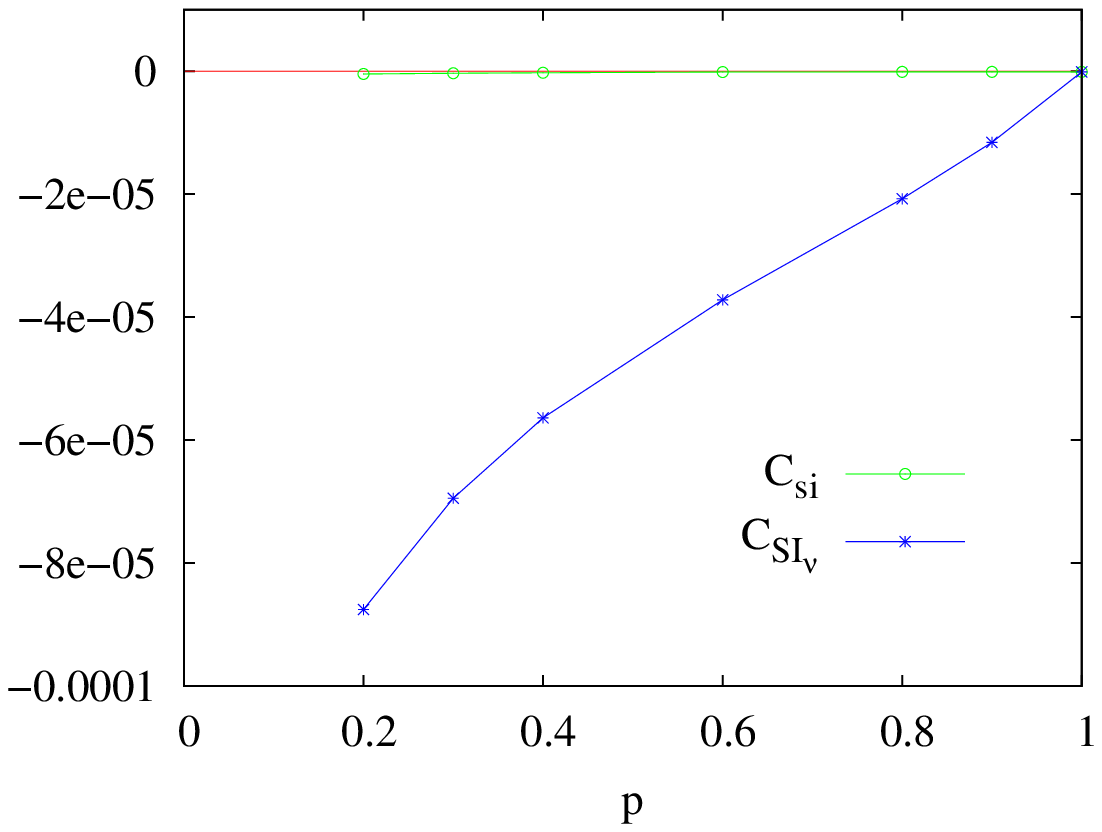}}
\subfigure[ ] {
\includegraphics[width=7.8cm,height=5.5cm]{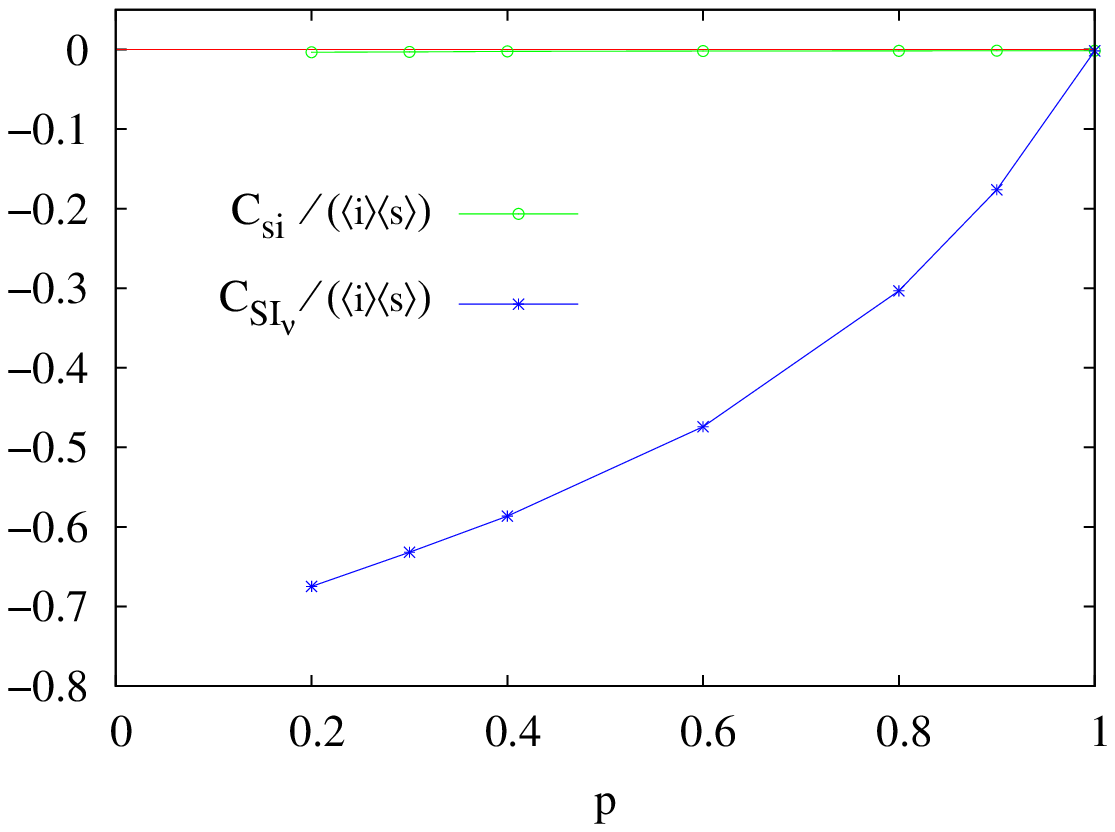}}
\caption{\label{figcorrel}
 (a)  Correlation coefficients $C_{s i}$ and $C_{S I_\nu}$ for $k=8$ networks
as a function of $p$. (b) The same, but correlations are divided
by the product \iav\ \sav.
}
\end{figure}
In Fig. \ref{figcorrel} we plot both correlations 
for the case $k=8$. 
While the magnitude of $C_{s i}$ is barely noticeable, 
$C_{S I_\nu}$ indicates that there is 
a strong reduction in the probability of having S-I pairs 
of neighbours as p decreases.
For $p=1$, \beff\ is almost equal to $\beta$ (the deterministic value) 
because the correction introduced by fluctuations is 
lower than 0.5\%. For $p<1$ local correlations introduced by the 
network are appreciable and cause the fall down in \beff .
The definition we take for \beff\ (Eq. \ref{defbeff})
and Eq. (\ref{ist}) imply that
\begin{equation}
\left< s \right> =\frac{\gamma + \mu}{\beta_{eff}}
\label{sbeff}
\end{equation} 
This relation holds for all the networks independently of 
$p$ and $k$. If we take \beff\ values from Fig. \ref{figbeff}
and compute $\left< s \right>$ through Eq. \ref{sbeff},  
we obtain a set of curves that collapse with those in 
Fig. \ref{figsi} (upper panel) validating numerically 
Eq. \ref{sbeff}.

In summary, 
the relations among stationary values and model parameters
for the SIR deterministic model hold exactly the same for
the averages at the QSS of the SIR stochastic model in all the networks 
considered 
if the rate of infective 
contacts, $\beta$, is replaced by the effective 
transmission rate, \beff . That is, it is possible to account
for all the network effects on stationary averages
by the rescaling of a single parameter.

\subsection{Dynamic behaviour and fluctations}
\begin{figure}[ht]
\subfigure[ ] {
\includegraphics[width=7.8cm,height=5.5cm]{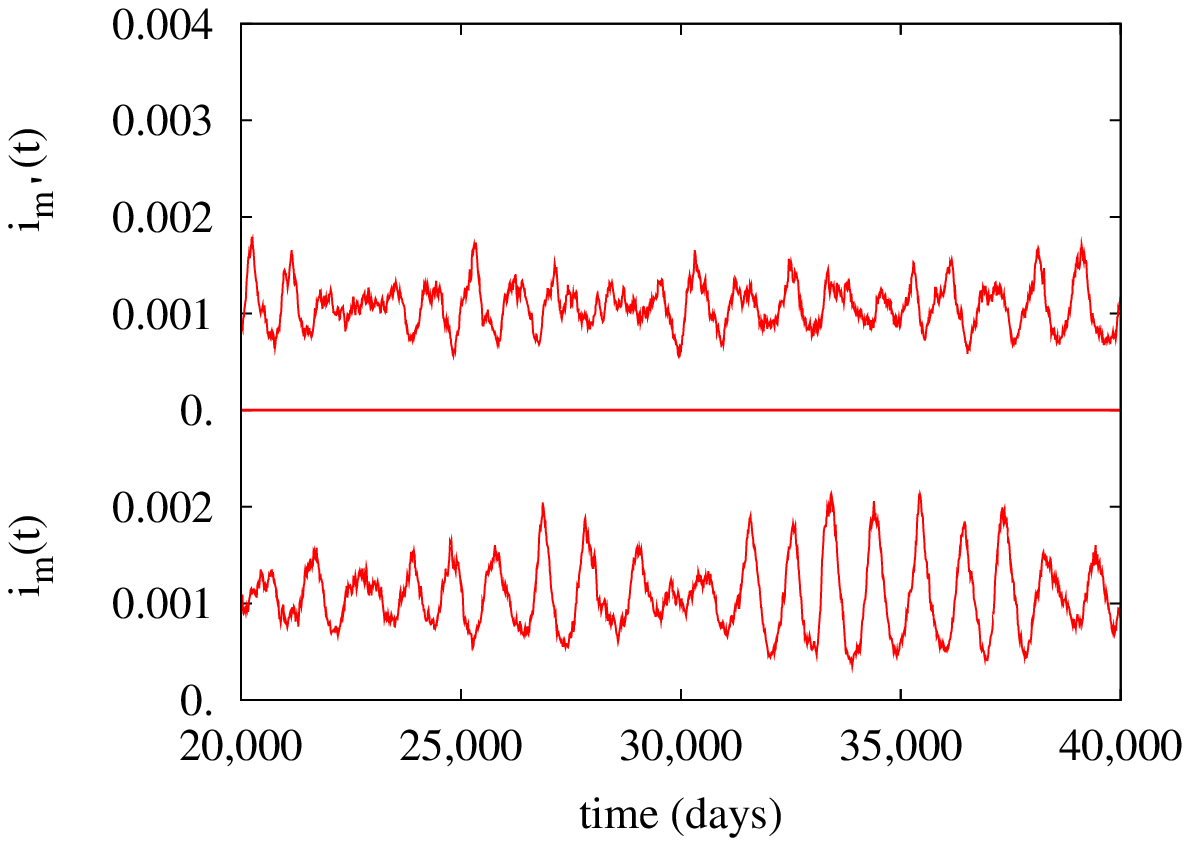}}
\subfigure[ ] {
\includegraphics[width=7.8cm,height=5.5cm]{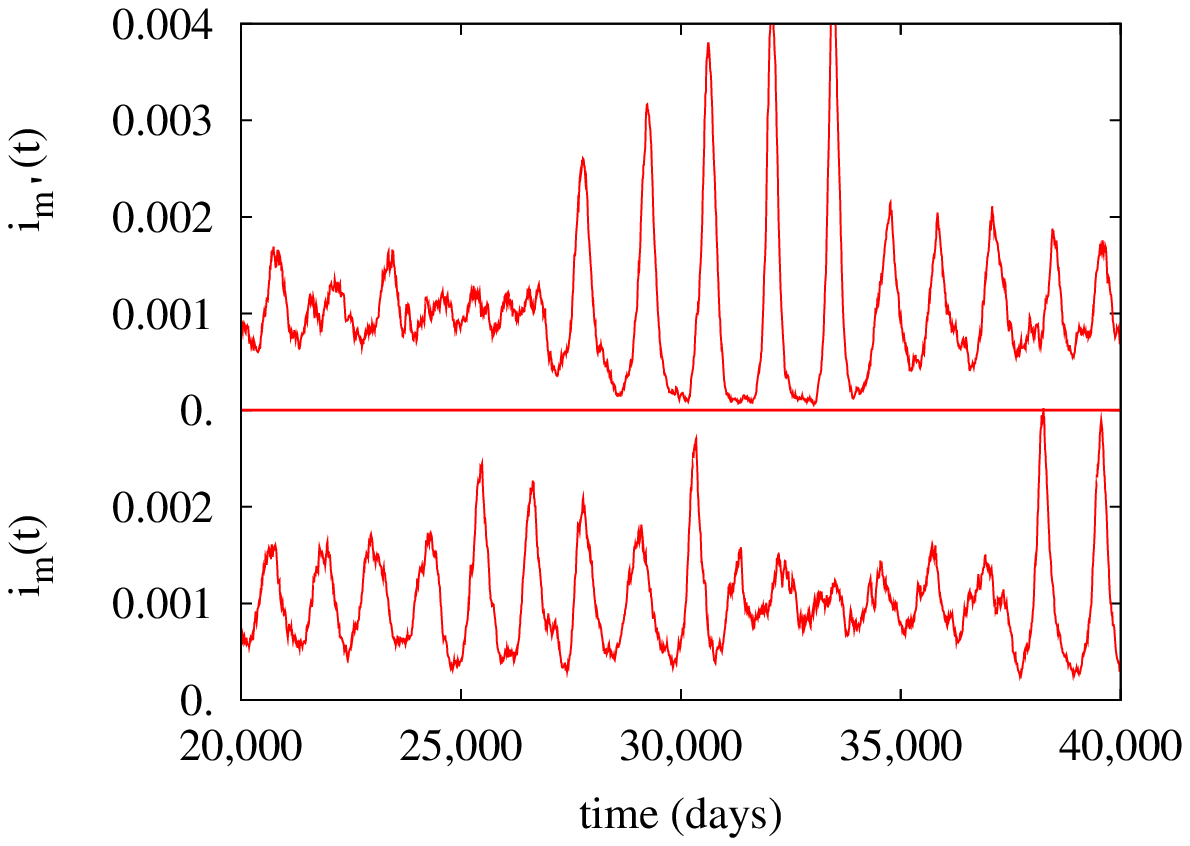}}
\caption{\label{figit} Time evolution of the fraction of infected
individuals in the QSS for two different samples
with $p$=1 (a) and $p$=0.3 (b).
}
\end{figure}

In this section we focus on the dynamic behaviour of the system 
in the quasi-stationary state.
We discuss the case $k$=8 since the other cases considered ($k$=4
and $k$=12) present similar qualitative behaviour.
In Fig. \ref{figit} we show the time evolution 
of the fraction of infected individuals for two samples corresponding
to  networks k=8, p=1, and k=8, p=0.3. 
Fluctuations are clearly larger for the case p=0.3 (Fig. \ref{figit}b) 
than for $p$=1 (\ref{figit}a). But, the amplitude of fluctuations 
changes considerably within the time evolution of a given sample. 
For example, for the case $p$=1 (sample $m$) in the 18 years between 
t=31,400 days and t=38,000 days (Fig. \ref{figit}a, lower panel) 
the amplitude of fluctuations is larger than for case $p$=0.3 (sample $m$)
in the 17 years between t=31,000 and t= 37,200 
(Fig. \ref{figit}b, lower panel).

\begin{figure}[ht]
\subfigure[ ] {
\includegraphics[width=7.8cm,height=5.5cm]{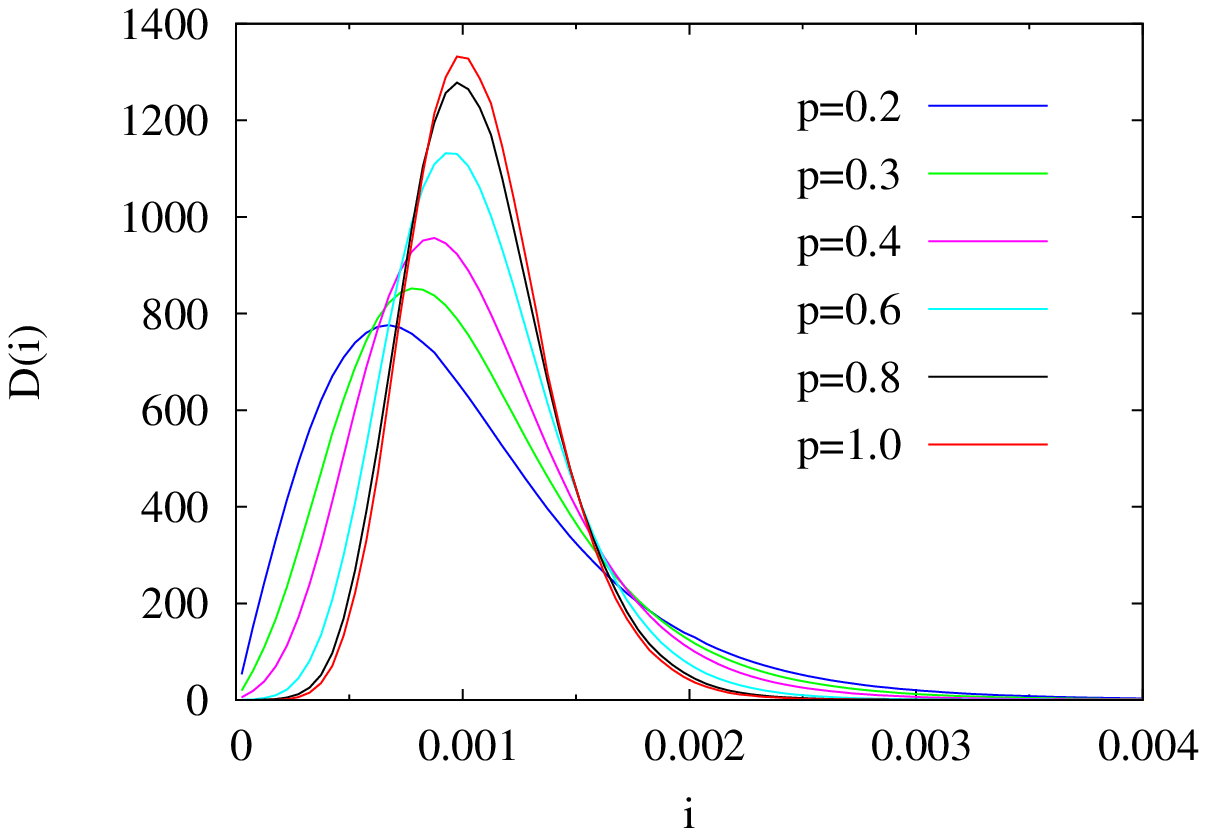}}
\subfigure[ ] {
\includegraphics[width=7.8cm,height=5.5cm]{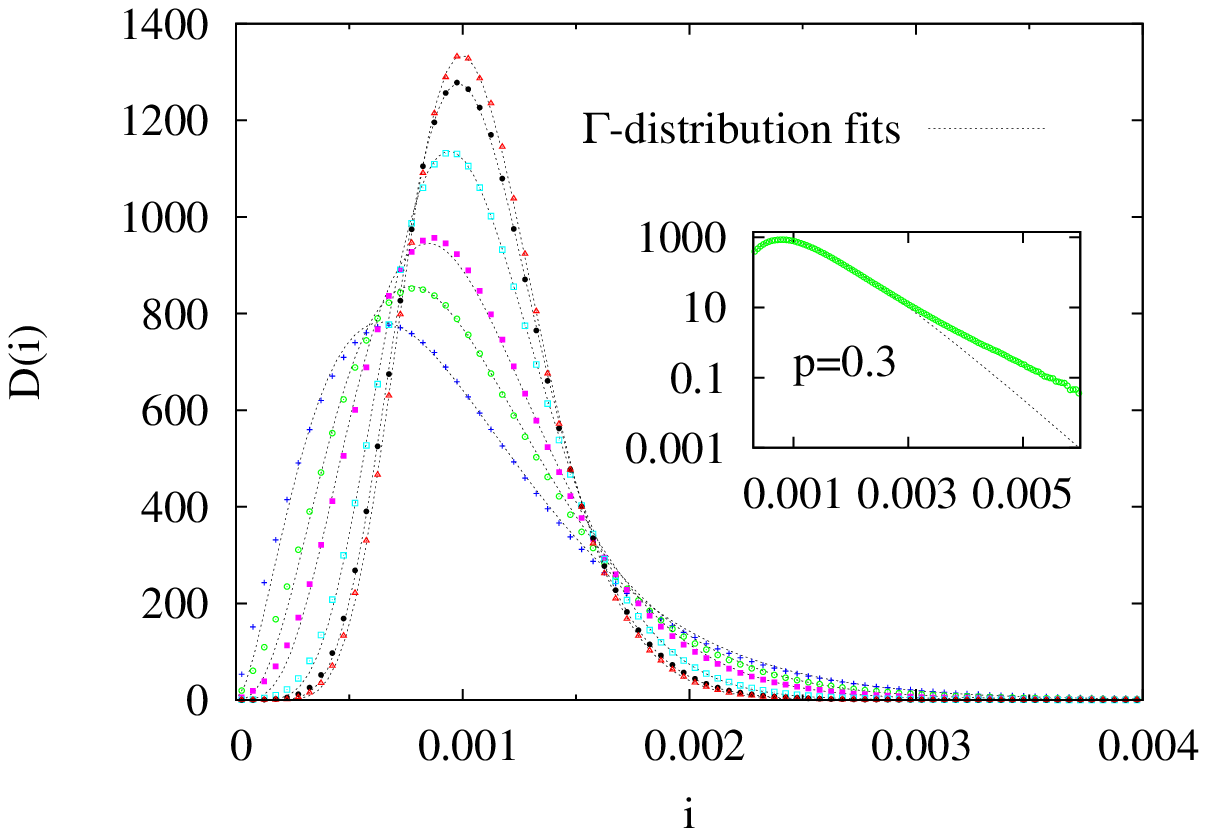}}
\subfigure[ ] {
\includegraphics[width=7.8cm,height=5.5cm]{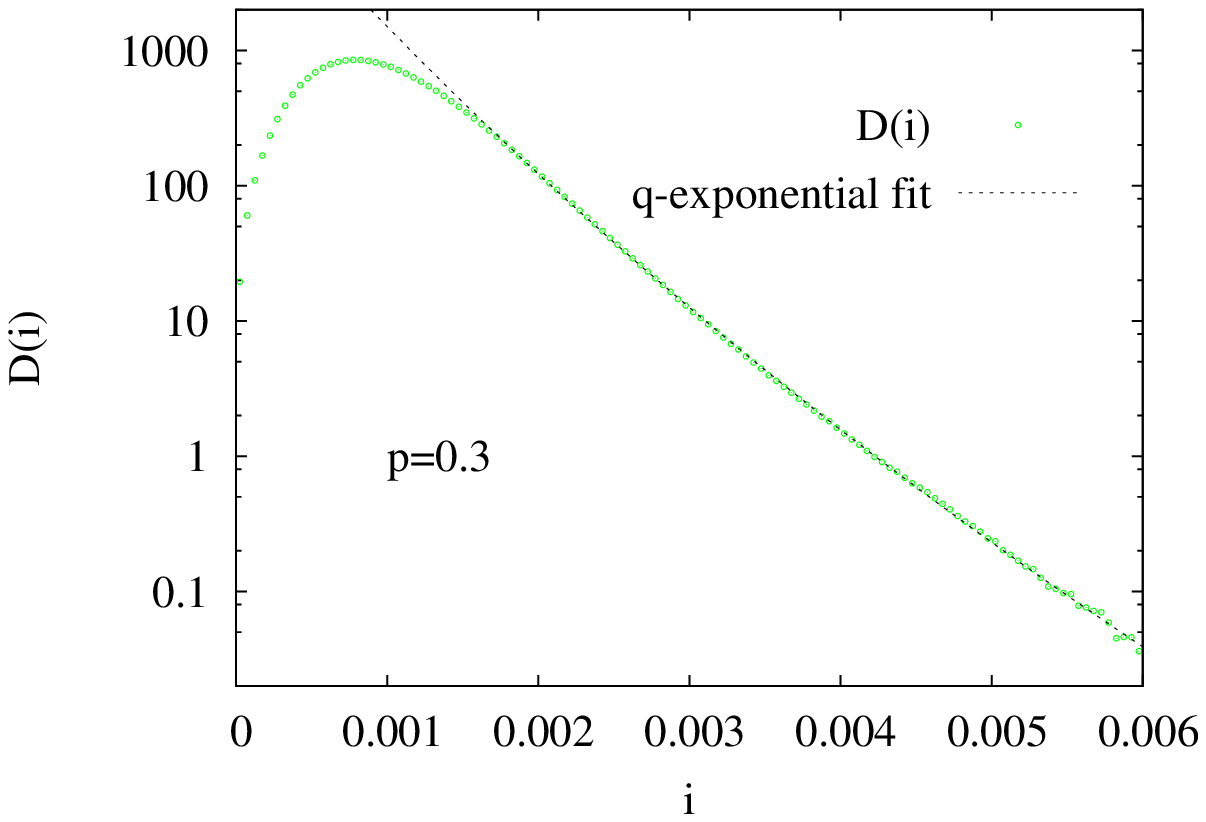}}
\subfigure[ ] {
\includegraphics[width=7.8cm,height=5.5cm]{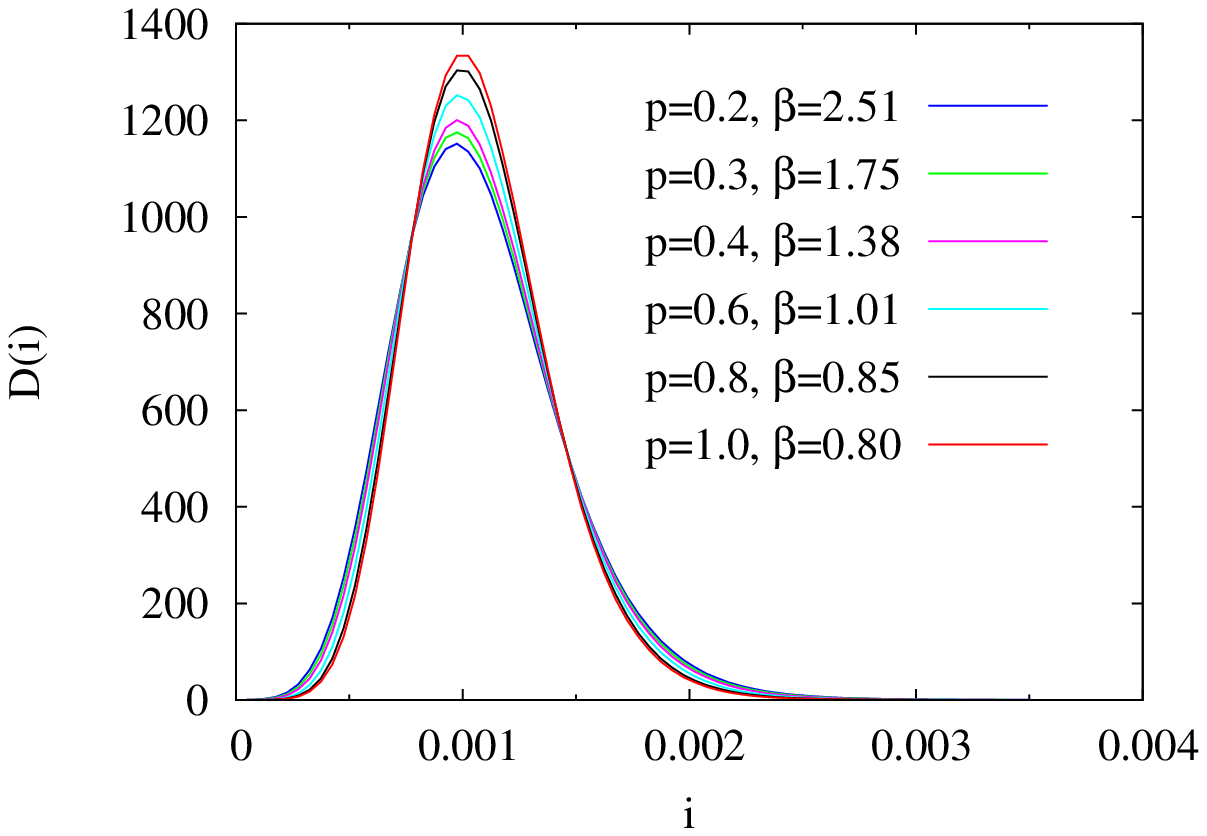}}
\caption{\label{figPi} PDF of the fraction of infected individuals, $D(i)$,
for networks with $k=8$. (a) $D(i)$ for different values of $p$ parameter.
To construct the histograms we use the $M$ trajectories
keeping an $i$ value every 10 days. This gives more than $2.10^7$ $i-$values
per histogram. (b) $\Gamma$-distribution fits (solid lines) to the
$D(i)$ of (a) (points). 
Inset: detail for case $p$=0.3 showing the
fit  underestimation of the $D(i)$ tail.
(c) $q$-exponential fit ($q$=1.04) to the $D(i)$ tail for case $p$=0.3. 
(d) $D(i)$ for networks with different $p$-values and 
taking different $\beta$-values in order
to obtain the same effective transmission rate: \beff=0.8. 
}
\end{figure}

With the aim of characterizing the fluctuations for the networks considered, 
we compute the  probability density functions (PDF), $D(i)$, 
that are histograms constructed from the instantaneous
fractions of infected individuals.
In Fig. \ref{figPi}a we show the PDF for networks with $k$=8 and different 
$p$ values. 
When $p$ decreases, distributions become more asymmetrical with long tails
and maxima shifted towards low $i$-values. The combination of both features
results in the small  variation of \iav\ with $p$ that we
observed in Fig. \ref{figsi}.
The long tails indicate the presence
of large peaks of infected individuals as $p$ decreases.
We found that PDF are well fitted by $\Gamma$-distributions
\begin{equation}
G(x)= \frac{1}{\Gamma(n) \lambda ^n} \ x ^{n-1} e^{-x / \lambda}
\end{equation}
where $\lambda$ and $n$ are the only two fitting parameters. 
The fits are compared with $D(i)$ histograms 
in Fig. \ref{figPi}b. The fits are good except for very low
and very high values of $i$. The fitted $\Gamma$-distributions
underestimate the $D(i)$ tails that fall more slowly than exponentials 
(see inset of Fig. \ref{figPi}b).
In fact, the  $D(i)$ tails are well fitted by $q$-exponentials,
$C  e_q (-\lambda i)$, where
\begin{equation}
e_q(x)= \left[ 1 + (1-q) x \right] ^{1/(1-q)}
\end{equation}
is an exponential in the limit $q \rightarrow 1$ \cite{qexp}.
In Fig. \ref{figPi}c the $D(i)$ tail for case $p=0.3$ is compared
with the $q$-exponential fit ($q$=1.04) showing an excellent agreement
for three decades of $D(i)$-values.

In summary, the study of $D(i)$ distributions reveals
a complex behaviour of the instantaneous values of
the fraction of infective individuals, $i$, for different networks
that is masked by the near constancy of \iav\ with $k$ and $p$ observed before.
The question that arises at this point 
is whether this complex behaviour is only due to a change in \beff.
In order to answer this question, we have performed
simulations of the system for different networks (different $p$ and $k$ values)
but changing the value of parameter $\beta$ in order to
obtain the same \beff\ for all cases.
In Fig. \ref{figPi}d we show the $D(i)$ distributions for different
networks with $k=8$ and the same value of \beff=0.8. 
The collapse of the curves for different $p$ values points out that 
the change in \beff\ was responsible for the 
main effects of the network on $D(i)$ 
that we observed in Fig. \ref{figPi}a.               
However, the collapse of the curves in Fig. \ref{figPi}d 
is not complete, they are a bit broader 
for lower $p$ indicating
that for the same \beff, a decrease in $p$ produces 
slightly larger fluctuations.
The change in the mean squared deviation, $\sigma$, 
and the related change in the peak height are the
only relevant differences among these curves
that don't show the large asymmetry of those of
Fig. \ref{figPi}a.

\begin{figure}[ht]
\subfigure[ ] {
\includegraphics[width=7.8cm,height=5.5cm]{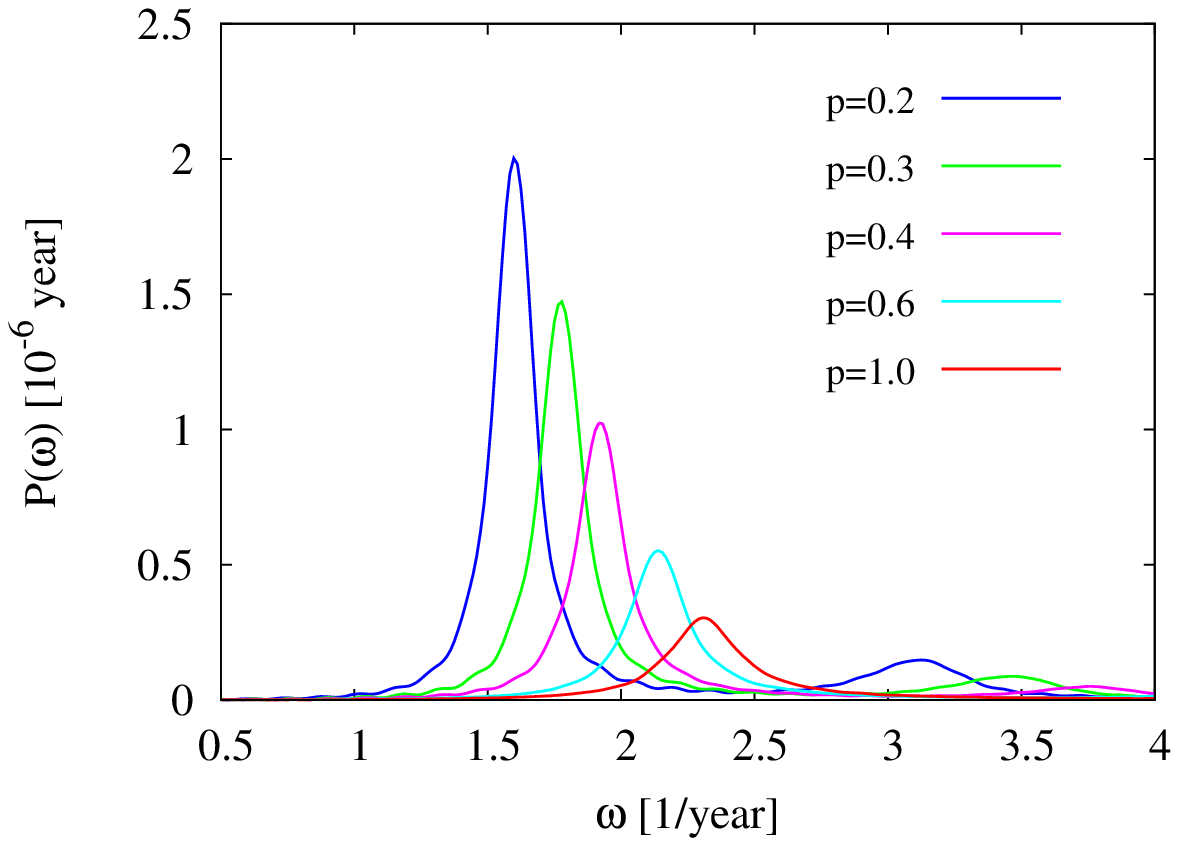}}
\subfigure[ ] {
\includegraphics[width=7.8cm,height=5.5cm]{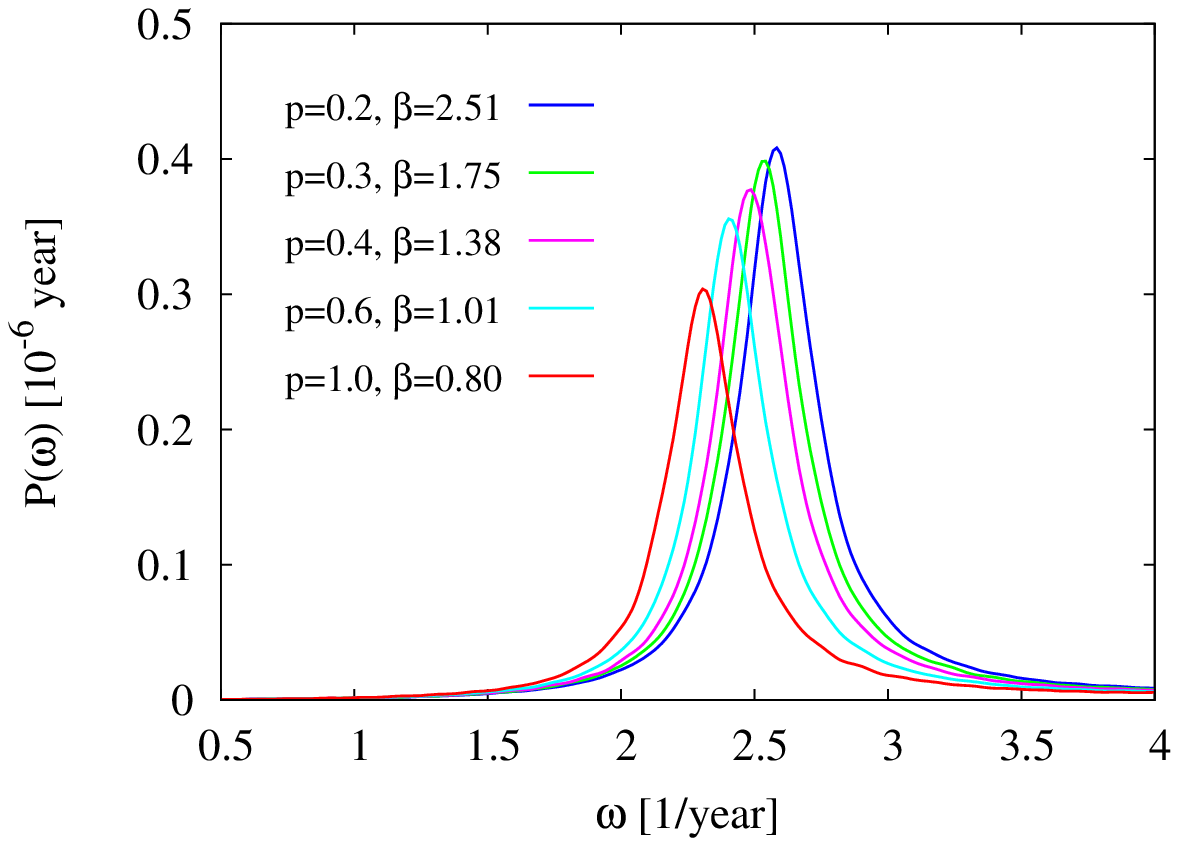}}
\caption{\label{figPw}(a)  Fluctuation power spectra of infective time
series, $P(\omega)$,
 for networks with $k=8$ and different $p$-values, 
(b) $P(\omega)$ for the same networks considered in (a)
but taking different $\beta$-values in order
to obtain the same effective transmission rate: \beff=0.8.
}
\end{figure}

To study the dynamical behaviour of the fraction of infective
individuals, 
we compute the self-correlation function
\begin{equation}
c(t)= \left< i(t') i(t'+t)  \right> - \left< i \right>^2
\label{selfcorrel}
\end{equation} 
which results independent of $t'$ in the $(t_a,t_b)$ interval.
By Fourier transforming $c(t)$, we obtain the power spectrum 
$P(\omega)=\int \cos(\omega t) c(t) dt$
for the different networks considered (Fig. \ref{figPw}a). 
As $p$ decreases, the peak of $P(\omega)$ shifts towards lower values of
$\omega$ and becomes more pronounced. The area under $P(\omega)$
is proportional to $\sigma ^2$ \cite{antitr}, indicating that 
the amplitude of fluctuations increases as $p$ decreases, as could also
be inferred from the increasing width of $D(i)$s (Fig. \ref{figPi}a).
This large  increase in $P(\omega)$ peaks when $p$ is reduced, 
as well as the presence of 
a secondary harmonic peak have already been observed
in a previous study of this system for other 
model parameters \cite{simoes}.
For the case $p=1$ an analytical expression that
approximates very well $P(\omega)$ has been derived \cite{alonso}.
In order to sense how many of the network effects on dynamic 
correlations can be 
 accounted for by the effective transmission rate, we compute 
$P(\omega)$ for different networks having the same 
\beff. The curves, shown  in Fig. \ref{figPw}b,
present a slight increase of the peak height as $p$ decreases.
The resulting increase in the area under $P(\omega)$
is consistent with the slight increase in width ($\sigma$) of $D(i)$
observed in Fig. \ref{figPi}d.
It can be seen that even
when \beff\ is kept fixed, 
there is a shift of the peaks towards greater frequencies 
as $p$ decreases. These effects depend on the specific
set of parameters taken for the SIR model. For example, in Ref.
\cite{simoes} the authors found 
   (for other values of model parameters)
 that when keeping \beff\ fixed,
the height of $P(\omega)$ peak is 
highly increased when $p$ is lowered, while the shift in frequency 
is barely noticeable (see Fig. 1c of Ref. \cite{simoes}). 

\section{Discussion}

Our study of the SIR model on a Watts-Strogatz-type network, for parameters
corresponding to pertussis in the pre-vaccine era, shows that network
structure strongly influences the disease dynamics. The increasing locality 
of the network (obtained by lowering $p$) decreases the 
disease transmission. This effect has already been observed by other 
authors \cite{verdasca,simoes} who attributed it to the clustering of
infected individuals produced by local correlations. 
In the present work we 
quantified this concept obtaining an explicit relation
between the effective transmission rate, \beff, and correlation coefficients 
between S and I individuals (Eqs. \ref{beffcorr} and \ref{coefcorr}).
An increase in  locality also drastically increases fluctuations
and the period between outbreaks, which  have been analyzed
characterizing the probability density functions 
of fraction of infected individuals, 
$D(i)$, and the power spectrum of the $i$-time series, $P(\omega)$.

We analyzed whether the network plays another role 
in the behaviour of the system at the quasi-stationary state 
{\it besides} the change in the
effective transmission rate of the disease. Our findings are summarized below:
\begin{itemize}
\item [-] The average fractions of susceptible, infected and recovered
 individuals at the QSS
 have the same values as  in the SIR deterministic model
if $\beta$ parameter is replaced by \beff\ of the corresponding network.
Therefore, for \sav\ and \iav,
the network structure may be ignored if a single parameter is changed properly.

\item [-] Concerning fluctuations and the time-correlated 
behaviour of the system, the situation
is highly dependent on the SIR-model parameters taken. In particular, 
if we consider different networks taking the appropiate value
for $\beta$ in order to obtain \beff=0.8, 
fluctuations in the fraction of infected individuals 
and their time correlations 
are very similar as those of the
stochastic SIR model ($p=1$) for all the networks considered
(Fig. \ref{figPi}d and Fig. \ref{figPw}b).
However, network effects are not completely absorbed by \beff\ in 
this case.
\end{itemize}
From these remarks we are able to answer 
the question we raised
in the introduction: If we want to describe pertussis transmission
in the pre-vaccine era, what is lost if we use a SIR stochastic model
with homogenous mixing instead of a SIR stochastic model on the network
with determined values of $k$, $p$ and $\beta$?
The answer is: very little. 
In order to parametrize both models, 
what can be taken from epidemiological data is the basic 
reproductive ratio $R_0$ that, for pertussis in the prevaccine era, 
is between 16 and 18 \cite{libroAM, AMjhyg94} and 
may be obtained from the average fraction of susceptibles, \sav, 
through the relation: $R_0$=1/\sav\ \cite{nokesanderson}. 
As for the studied systems 
\sav=\beff /$(\gamma+\mu)$, 
epidemiological data fixes \beff\ (not $\beta$) around 0.8.
If the SIR stochastic model ($p$=1)
is used, the frequency of outbreaks would be a bit underestimated
with respect to the prediction of a network with $p<1$ (Fig. \ref{figPw}b).
But it would be very difficult to infer 
the proper $k-p-\beta$ combination
from measurable quantities. 
For example, it is very unlikely that with a power spectrum
constructed from a time series
of 20 or 30 years
it will be possible to
choose  among one of the curves of Fig. \ref{figPw}b. 
In fact, given
the high heterogeneity in 
the time series obtained as output of the model (Fig. \ref{figit})
care should be taken when analyzing real data. If the SIR stochastic
model is
proposed to describe the dynamics of pertussis in the pre-vaccine
era and in cities of the size considered in the present work,
the differences observed in the incidence time
series in different countries might be  due to the
heterogeneity of the time profile itself
and not to differences in the epidemiological conditions
at each place \cite{incidences}. 
It would be of great interest to know whether these conclusions
may be extended to more complicated and realistic models.
In particular, epidemiological and laboratory studies suggest that
 immunity acquired by pertussis infection is not lifelong 
(see ref. \cite{wendelboe} and references therein) and more 
realistic models of pertussis transmission include 
compartments that account for waning immunity
\cite{Hethcote99,vanboven,nos-adol-boost}.

We find it appropriate to emphasize that the results and conclusions 
obtained in this work hold: 
   a) for the SIR stochastic model in the dynamical networks 
      with local and global contacts as described in section 2.2, 
   b) for the set of parameters used that corresponds to pertussis disease 
      in the pre-vaccine era and 
   c) for the quasi-stationary state of the system, as empirically defined 
      in section 3.1. 
Extrapolations to other problems that share {\it only some} features 
with the ones treated in the present work are not straightforward 
and require caution. 
  We expect relations (14), (16) and (18) concerning stationary values 
to hold independently of the parameters used provided that the 
quasi-stationary state has been established. 
But concerning fluctuations we have not obtained general relations and 
we do not expect that conclusions from the results presented here 
will be valid for systems with parameters representing other 
infectious diseases. 
There are several open questions related to the present work 
that could be addressed in future research.
In particular, it would be interesting 
to know whether our description of the system at the quasi-stationary 
state with a beta-rescaled stochastic SIR model could be 
extended to other model parameters. 
Moreover, it would be interesting to study whether our description 
also holds in the approach to the steady state. 
The problem of constructing mean field approximations to describe 
the approach and behavior of epidemic systems in the steady state has been 
studied for the SIRS model on static two-dimensional Watts Strogatz 
networks by Roy {\it et al}. \cite{roypascual}.
Concerning the comparison between static 
and dynamical networks, in ref. \cite{zanette} Zanette studied the 
dynamics of rumor propagation with an SIR-type model (without mortality) 
in standard (1D) Watts Strogatz network and in its dynamical version 
defined as in the present work. They found that the qualitative behavior 
of propagation is the same in both networks but that the effectiveness 
of propagation is considerably higher in the dynamical one. 
This sort of comparison would be interesting for infectious disease 
transmission modeling because both effects of the $p$-parameter 
(randomness and globality) could be analyzed separately. 
While in static networks $p$ defines the average fraction of fixed random 
contacts, in dynamical networks $p$ also measures the degree of globality 
of social contacts (contacts with any individual in the population).

Finally, it has to be mentioned, that our definition of 
\beff\ (Eq. \ref{defbeff}) differs
from that of Refs. \cite{verdasca} and \cite{simoes} where 
\beff= $\left< a_{inf} / (isN) \right>$.  
In any case, both definitions throw
very similar results for all the networks considered in this work 
(the relative difference is always below 0.3\%).

\addvspace{20pt}
\section*{Acknowledgments}
We aknowledge Alberto Maltz for fruitful discussions.
This work was supported by Agencia Nacional
de Promoci\'on Cient\'{\i}fica y Tecnol\'ogica-ANCPyT,
and Consejo Nacional de Investigaciones
Cient\'{\i}ficas y Tecnol\'ogicas-CONICET
(Argentina). G.F. is member of the Scientific Career of CONICET.

\addvspace{20pt}
\section*{References}

\addvspace{20pt}


\begin{thebibliography}{00}
\bibitem{antitr} The inverse Fourier transform of $P(\omega)$
at the origin is: $c(0)=2/\pi \int P(\omega) d\omega$, and
$c(t)$ at the origin is $\sigma ^2$ (Eq. \ref{selfcorrel}).
\bibitem{incidences} We here refer to the incidence (the
infections produced per unit of time) because it is the
magnitude available from epidemiological data. 
Even though in the present work 
the fraction of infected individuals, $i(t)$, was analyzed, we
checked that
the incidence computed with the model shows the same dynamic behaviour
as that of  $i(t)$
producing curves very similar to those of Figs. \ref{figit},
\ref{figPi} and \ref{figPw}.
\end{thebibliography}
\end{document}